\documentclass[]{JHEP}

\usepackage{amssymb}
\usepackage{epsfig}

\preprint{DAMTP--2001--45\\
LBNL--48377}

\title{Branes on  charged dilatonic backgrounds:
self-tuning, Lorentz violations and cosmology}

\author{Christophe Grojean\\
	Department of Physics University of California\\
	Berkeley, CA 94720 USA, and\\
	Theoretical Physics Group, LBL\\
	Berkeley CA 94720, USA\\
	E-mail: \email{CMGrojean@lbl.gov}}

\author{Fernando Quevedo and Ivonne Zavala C.\\
	Centre for Mathematical Sciences, DAMTP, University of Cambridge,\\
	Cambridge CB3 0WA UK\\
	E-mail: \email{F.Quevedo@damtp.cam.ac.uk}, 
		\email{E.I.Zavala-Carrasco@damtp.cam.ac.uk}}

\author{Gianmassimo Tasinato\\
	SISSA, Via Beirut 2-4, 34013  Trieste, and\\ 
	INFN, Sezione di Trieste, Italy\\
	E-mail: \email{tasinato@he.sissa.it}}

\abstract{We construct an $n+q+2$ dimensional background that has
dilatonic $q$-brane singularities and that is charged under an
antisymmetric tensor field, the background spacetime being maximally
symmetric in $n$-dimensions with constant curvature $k=0,\pm1$.  For
$k=1$ the bulk solutions correspond to black $q$-branes.  For $k=0,-1$
the geometry resembles the `white hole' region of the
Reissner-N\"ordstrom solution with a past Cauchy horizon. The metric
between the (timelike) singularity and the horizon is static whereas
beyond the horizon it is cosmological.  In the particular case of
$q=0$, we study the motion of a codimension one $n$-brane in these
charged dilatonic backgrounds that interpolate between the original
scalar self-tuning and the black hole geometry and provide a way to
avoid the naked singularity problem and/or the need of having exotic
matter on the brane.  These backgrounds are asymmetrically warped and
so break 4D Lorentz symmetry in a way that is safe for particle
physics but may lead to faster than light propagation in the
gravitational~sector.}


\begin{document}

\section{Introduction}

The brane-world scenario is providing new ideas to approach old
questions such as the hierarchy problem, the cosmological constant
problem and early universe
cosmology~\cite{early}--\cite{ccXdim}.  The simple idea
that our Universe is a brane trapped in a higher dimensional space
allows for a great amount of possible realizations depending on the
distribution of matter on the brane and the bulk as well as their
relative dimensionality.

 As usual, simplicity has been the main guideline when considering
explicit realizations of the brane world.  Most discussions in the
current literature refer to 3-branes inside a five dimensional bulk
with only gravity propagating in the bulk. Adding extra fields to
solve some of the problems such as radius stabilization and to
ameliorate the cosmological constant problem has also been
considered~\cite{ccXdim}--\cite{CEG}.  But in principle there is a
great degree of arbitrariness and in order to go beyond the simplest
realizations we need to have a general guideline.

Clearly the best motivated brane world scenarios are those that can
naturally be obtained from string theory~\cite{fromstrings}. There are
actually at present few explicit realizations of quasi-realistic brane
world models derived from string theory~\cite{models}. We can try to
extract the general properties of those models to incorporate them on
a particular framework to approach different phenomenological and
cosmological properties of these scenarios.

\pagebreak[3]

Following this guideline we will consider in this note a system
consisting of a $q$-brane\footnote{We refrain from using the standard
terminology $p$-brane to avoid confusion with the pressure $p$ in the
following sections.}  singularity in a $d=n+q+2$ dimensional bulk with
gravity, dilaton and antisymmetric tensor fields.  We find explicit
solutions for the field equations for which the {$n$-dimensional}
slices of the spacetime have constant curvature $n(n-1)k$, $k=1,0,-1$
(for a related discussion see for instance~\cite{mirjam2}).  A
motivation for the study of these geometries is their potential
application to brane cosmology. One of the most interesting results
emerging in this field has been the realization, through Birkhoff's
theorem, that an additional $n$-brane\footnote{For clarity we mention
the distinction between the two branes involved in our construction:
the $q$-brane that is electrically charged under a $q+1$ form and an
$n$-brane that is coupled to the bulk through gravity only. Until
section 3, we will keep $q$ and $n$ general and we will restrict to
$q=0$ afterwards.} carrying ordinary matter and gravitationally
coupled to the bulk, when moving in a static black hole background,
actually feels a time dependent cosmology~\cite{kraus,KeKi,BCG}.
Therefore it is clear that the solutions presented in this work
provide interesting brane cosmology backgrounds even in the regions
where they are static.
  
\looseness=-1The causal structure of the spacetime depends on the topology of the
dimensions parallel to the external $n$-brane.  The $k=1$ case has
been studied in the past and it corresponds to black
$q$-branes~\cite{HS}, with $n+q+2=10$. The global geometry consists on
an asymptotically flat spacetime with a horizon and two
singularities.\footnote{Note however, that in string theory, black
holes will carry several charges, and then, the extra singularity can
be stabilized~\cite{miriam}.}  The $k=0,-1$ solutions we find are not
black $q$-branes. They have an interesting global structure with a
horizon and only one singularity at the origin. The region between the
singularity and the horizon is static, unlike the standard black hole
case. Beyond the horizon it becomes time dependent, therefore
corresponding to a cosmological solution for which there is no
singularity at any surface of constant time.

For the $q=0$ case, the motion of the external $n$-brane (a
codimension one brane) can be studied easily using the usual Israel
junction equations.  We mostly concentrate in the regions of the bulk
spacetime which are static and find the possible places where the
brane can be located, cutting the space in the transverse dimension
such that a $\mathbb{Z}_2$ symmetry around the brane can be imposed
and a finite Planck scale in four dimensions is guaranteed.

\looseness=-1We find that for $k=1$ the $n$-brane can be located in the region of
the bulk spacetime which is outside of the singularities and the
horizon and therefore the extra dimension can be naturally restricted
to be the region between the black hole horizon and the location of
the brane, avoiding naked singularities completely. The matter in the
brane is not exotic because the value of the parameter $\omega$
relating the pressure and energy density on the brane ${p}=\omega
\rho$ lies in the physically allowed region $0>\omega>-1$.

For $k=-1,0$ the situation is different. The static region lies
between the singularity and the horizon, therefore there are two
possibilities. The space can be taken between the brane and the
horizon or between the brane and the singularity. In the first case it
turns out that the energy density of the brane has to be negative and
in the second it is positive. In both cases $\omega$ can take
physically allowed values.

We will consider in more detail the simplest system of a 3-brane in a
five-di\-mensional bulk with dilaton and a two-index antisymmetric
tensor field $B_{\mu\nu} $.  Since~in five-dimensions the
antisymmetric tensor is dual to a vector field $\partial_\mu A_\nu =
\epsilon_{\mu\nu\rho\sigma\tau}\partial^\rho B^{\sigma\tau}$, this
situation is equivalent to consider a gauge field $A_\mu $ instead of
$B_{\mu\nu} $. Interestingly enough the introduction of a dilaton
field has been studied to ameliorate the cosmological constant
problem~\cite{ADKS,KSS} and more recently a similar proposal was made
regarding the introduction of a gauge field~\cite{CEG}. It is then
natural to consider both fields together.

This generalization improves the situation with only one of the fields
in several respects regarding the self-tuning of the cosmological
constant.

\begin{itemize}
\item For vanishing gauge fields we reobtain the solution
of~\cite{ADKS,KSS} as a particular case of our $k=0$ solution. However
our solutions also include possibilities not considered
in~\cite{ADKS,KSS} since we consider asymmetrically warped geometries
which are not 4D Poincar\'e invariant from the bulk point of view but
are so at each location of the brane, by suitably redefining the speed
of light.  These cases share the property with~\cite{ADKS,KSS}\ that
there are always naked singularities and require a fine-tuning of the
dilaton couplings and the parameter $\omega$ defining the equation of
state of the matter on the brane.

\item For non vanishing gauge fields and $k=0$, $\omega$ is still a
constant related to the dilaton couplings.  Indeed, contrary to the
model~\cite{CEG} with a pure Reissner-N\"ordstrom black hole in the
bulk, the presence of the scalar field requires a fine-tuning of the
dilaton couplings as in the original scalar self-tuning
models~\cite{ADKS,KSS}.  The improvement on the self-tuning relies on
the fact that, as mentioned above, the naked singularities can be
avoided as long as the energy density on the brane is negative. Also
$\omega$ can lie in the physically allowed region.

\end{itemize}

Since our solutions correspond to asymmetrically warped metrics they
may induce explicit violations of Lorentz invariance on the brane by
having gravitational waves moving faster through the bulk than through
the brane. We study the variation of the speed of light with the
location of the brane and find that gravitational waves move faster
through the bulk only in the cases with naked singularities.
 
The organization of the paper is as follows. We present in
section~\ref{sec2} the general solution of the bulk equations of
motion for gravity coupled to the dilaton field and an antisymmetric
tensor of rank $q+2$, for the cases $k=0,\pm 1$, generalizing the
results of~\cite{HS}\ (see also~\cite{GM}).  In the next section we
introduce the external $n$-brane with matter in a perfect fluid for
the particular case of codimension one brane. We obtain the junction
conditions and then obtain the geometry defined by the position of the
$n$-brane in the bulk background. We specialise in section~\ref{sec4}
to the five-dimensional case discussing in detail the issues of
self-tuning and violation of Lorentz invariance in these
backgrounds. Finally we present our conclusions and discuss some
issues related to the cosmological implications of the bulk solutions.

\section{General charged dilatonic $q$-brane background}\label{sec2}

In~\cite{kraus,KeKi} it was found that the cosmology of a brane inside
a higher dimensional bulk spacetime can be studied by considering a
\emph{static} bulk geometry with the same spatial structure, i.e., a
spacetime with a constant curvature, maximally symmetric subspace of
the same dimensionality as the brane. For an observer on the brane,
the movement of the brane in the static geometry becomes an evolving
brane Universe.  This is a direct consequence of Birkhoff's theorem in
more than 4 dimensions. In~\cite{BCG} it was explicitly shown how to
map the cosmological and static metrics for the case of a
five-dimensional bulk without extra matter fields.  Similar results
hold when there are background gauge fields~\cite{CEG}.  Notice that
this is not always the case, we may have some cosmological solutions
that may not be mapped to static ones in more general cases when other
fields are included. In particular, in presence of a scalar field,
Birkhoff's theorem does not hold~\cite{bgqtz}. Nevertheless, it is
still possible to find some background solutions that remain static
or, as we will see, that depend on the time coordinate only. Even if
these solutions are no longer the most general solutions in the bulk,
they are interesting on their own and we will focus on their study in
this paper.

In this section, we generalize the electric black $q$-brane solution
studied by Horowitz and Strominger in~\cite{HS} for positive and
constant spatial curvature ($k=1$) to the case of arbitrary dimensions
and arbitrary spatial curvature $k=0,\pm 1$.\footnote{The magnetically
charged case is straightforward.}

We consider the coupling of gravity to a dilaton field and an
antisymmetric tensor with the following action in the Einstein
frame:\footnote{ Our conventions correspond to a mostly positive
lorentzian signature $(-+\ldots +)$ and the definition of the
curvature in terms of the metric is such that a Euclidean sphere has
positive curvature. Bulk indices will be denoted by Greek indices
($\mu,\nu\ldots$) and brane indices by Latin indices
($a,b\ldots$). The Einstein tensor in the bulk will be denoted by
$G_{\mu\nu}=R_{\mu\nu}-\frac{1}{2}R g_{\mu\nu}$.}
\begin{equation}
\label{generalaction}
S= \int d^{n+q+2}x \sqrt{g} \left( \alpha R - \lambda (\partial
\phi)^2 - \eta e^{-\sigma \phi} F_{q+2}^2 \right),
\end{equation}
where $\phi$ is the dilaton field, $F$ is a field strength $(q+2)$
form. We have left the couplings $\alpha, \lambda, \eta, \sigma$
arbitrary.  Varying the action (\ref{generalaction}) yields the
following equations of motion:
\begin{eqnarray}
\alpha G_{\mu\nu} &=& -\frac{1}{2} \lambda(\nabla\phi)^2\,g_{\mu\nu} +
\lambda\, \nabla_{\mu}\phi\, \nabla_{\nu}\phi + 
\nonumber\\&&
+\eta\,e^{-\sigma\phi}
\left( (q+2) F_\mu{}^{\lambda_1\ldots\lambda_{q+1}}
F_{\nu\lambda_1\ldots\lambda_{q+1}} - \frac{1}{2}g_{\mu\nu} F^2
\right),\qquad
\label{einstein}\\
2\,\lambda\,\nabla^2 \phi &=&-\sigma\,\eta \, e^{-\sigma\phi} F^2\,,
\label{dilaton1}\\
\nabla_{\mu_1} \left( e^{-\sigma\phi}
F^{\mu_1\mu_2\dots\mu_{q+2}}\right) &=& 0\,.
\end{eqnarray}
We are looking for solution of this system for which the spacetime is
a (warped) product of a $q$ dimensional space giving the
dimensionality of the $q$-brane singularity ($q=0$ is simply a black
hole) and an $n+2$ dimensional spacetime where the $n$-dimensional
slices correspond to spaces of constant curvature.  The electrically
charged solution is given by
\begin{eqnarray}
ds^2 &=& h_-^{\frac{-4 \lambda q (n-1)^2 b}{\alpha n (n+q) \Sigma^2}}
\left( -h_+h_-^{1-(n-1) b}dt^2 +h_+^{-1}h_-^{-1+b} dr^2 +r^2h_-^{b}
dx^2_{n,k} \right)+
\nonumber\\&&
+\, h_-^{\frac{4 \lambda  (n-1)^2 b}{\alpha  (n+q) \Sigma^2}} dy^2_{q}\,,
\label{eq:metric}\\
\phi&=&\frac{(n-1) \sigma b}{\Sigma^2} \ln h_-\,,
\label{eq:dil}\\
F_{try_1\ldots y_q} &=&\frac{Q \epsilon_{y_1\ldots y_q} }{r^{n}}\,,
\qquad  \epsilon_{y_1\ldots y_q}=\pm 1\,,
\label{eq:F}
\end{eqnarray}
where $dx^2_{n,k}$ is an $n$-dimensional spatial maximally symmetric
metric of constant curvature $n(n-1)k$, $k=0,\pm 1$.\footnote{Notice
that in general, one can add a constant to the dilaton field solution,
$\phi \to \phi(r) + 2 \phi_0$, by redefining the form as $F \to F
e^{\sigma \phi_{0}}$.} The harmonic functions, $h_\pm$, depend on two
constants of integration, $r_\pm$, and are given by:\footnote{Another
solution has been found (in the special case of $n=3$ and $q=0$) in
\cite{BF} and it is given by $h_+ = k-(r_+/r)^{n-1}$ and $h_- = 1-
(r_- / r)^{n-1}$.}
\begin{equation}
\label{eq:h+-def1}
h_+ = 1-\left(\frac{r_+}{r}\right)^{n-1}\,, \qquad 
h_- = k- \left(\frac{r_- }{ r}\right)^{n-1}\,.
\end{equation}
We have defined the quantities $\Sigma$ and $b$, in terms of the
different parameters of the action, by the following expressions
\begin{eqnarray}
\Sigma^2 &=& \sigma^2 + \frac{4 \lambda}{\alpha} \frac{q(n-1)^2}{n(n+q)}\,,
\label{eq:Sigma}\\
b &=& \frac{2\alpha n \Sigma^2}{(n-1)(\alpha n \Sigma^2+ 4(n-1)\lambda)}\,.\qquad
\label{eq:b}
\end{eqnarray}
The electric charge, $Q$, of this background is related to the two
constants of integration by
\begin{equation}
\label{eq:Q}
Q^2 = \frac{4n(n-1)^2 \alpha\, \lambda
(r_+r_-)^{n-1}}{(q+2)!\,\eta\,(\alpha n \Sigma^2 +
4(n-1)\,\lambda)}\,,
\end{equation}
while another combination of the constants of integration will be
interpreted as the ``mass'' of the background.  The solution with a
vanishing electric charge, which will be of physical relevance
concerning the problem of the cosmological constant, will be presented
separately in the section~\ref{sec:selftuning}.

Even if formally satisfying the equations of motion, this solution
does not make sense physically in the regions where $h_-$ becomes
negative, i.e., $r<r_-$ for $k=1$ and anywhere for $k=0,-1$, because
$h_-$ appears in the solution with non-integer powers, contrary to the
case of a simple Schwarzschild or Reissner-N\"ordstrom black
hole. However, it is easy to construct a new solution that overcomes
this problem:\footnote{ Notice also that it would have been possible
to flip both signs in front of $(r_\pm/r)^{n-1}$ in $h_\pm$ while
keeping a real electric charge, however the origin $r=0$ would be a
naked singularity.}  the solution will still be given by the
expressions (\ref{eq:metric})--(\ref{eq:F}) with the quantities
$\Sigma, b, Q$ still related to the parameters of the action by
(\ref{eq:Sigma})--(\ref{eq:Q}) but the definitions (\ref{eq:h+-def1})
of the functions $h_\pm$ has to be replaced by
\begin{equation}
\label{eq:h+-def2}
h_+(r) = s(r) \left(1-\left(\frac{r_+}{r}\right)^{n-1} \right),
\qquad
h_-(r) = \left|k-\left(\frac{r_-}{r}\right)^{n-1}\right|,
\end{equation}
where
\begin{equation}
s(r)={\rm sgn} \left(k-\left(\frac{r_{-}}{r}\right)^{n-1}\right).
\end{equation}

Now, it is important to observe some of the geometrical
characteristics of this solution. First of all, it is not hard to find
that for all $k$, $r_+$ is always a horizon, as usual. The other
properties depend on the value of the constant curvature $k$:
\begin{itemize}
\item \textbf{\textit{k=1:}} By computing the scalar curvature, it is
possible to realize that $r=r_-$ is a scalar singularity for an
arbitrary value of $b$.\footnote{For $b=0$, $r=r_-$ is actually
another horizon as in the Reissner-N\"ordstrom solution.}  The
background is asymptotically flat and corresponds (for $d=10$),
written in the Einstein frame, to the black $q$-brane solution
constructed by Horowitz and Strominger~\cite{HS}.
\item \textbf{\textit{k=0} and \textit{k=--1:}} It is clear from the
expressions for $h_-$ above, that $r=r_-$ is just a regular point,
whereas $r=r_+$ remains a horizon. Furthermore, the coordinate $r$
becomes timelike in the region $r>r_{+}$ but remains spacelike for
$r<r_+$, exactly the opposite of Schwarzschild black hole.  The
$q$-dimensional singularity at $r=0$ is then timelike. Therefore we
are only interested on the static region of spacetime, and we will
locate the external $n$-brane on the region $0<r<r_+$. However we
would like to point out here that the region $r>r_+$ is interesting
{per se} for cosmology. In this region $r$ becomes the time coordinate
and the horizon $r_+$ is a past Cauchy horizon for this cosmological
solution.  Unlike the standard cosmological singularity $r=0$ becomes
a timelike singularity behind the horizon, resembling the `white hole'
region of the Reissner-N\"ordstrom solution, but having a single
horizon instead of two. Unlike the Schwarzschild solution, since the
singularity is timelike it may be avoided by a future directed
timelike curve in the region beyond the horizon.
\end{itemize}
We show in  figures~\ref{fig1} and~\ref{fig2} the corresponding Penrose
diagrams for these geometries illustrating the relevant regions.

\EPSFIGURE[t]{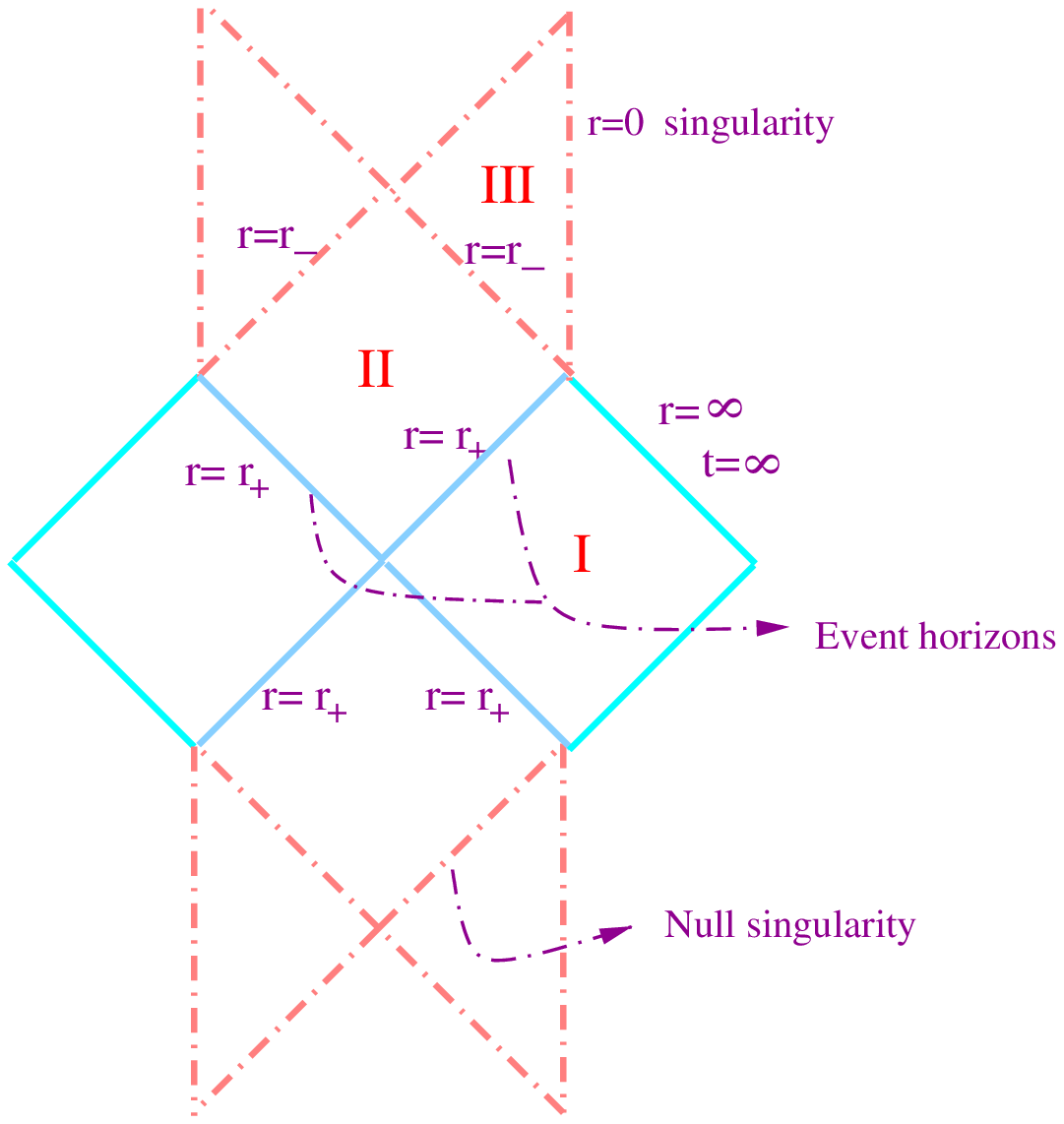, width=.6\textwidth}{Penrose diagram for the $k=1$ dilatonic
Reissner-Nordstr\"om black brane.\label{fig1}}

\EPSFIGURE[t]{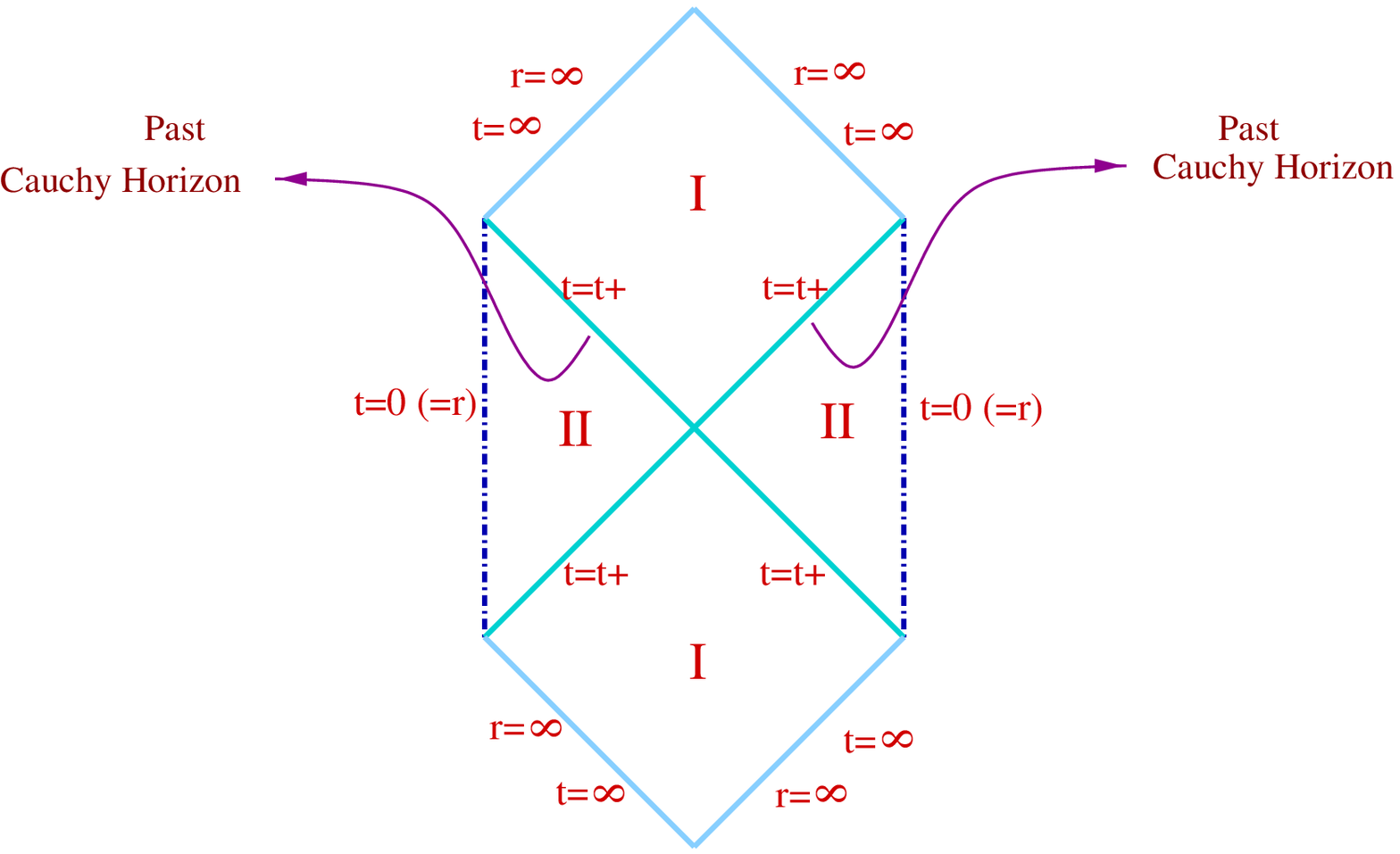, width=.75\textwidth}{Penrose diagram for the $k=0,-1$
case. This diagram is very similar to the Schwarzschild black hole
(rotated by $\pi/2$), but now region $I$ is not static, but
cosmological with a past Cauchy horizon and region $II$ is static. We
put the brane in region $II$.\label{fig2}}

A detailed study of the external $n$-brane inside the cosmological
regions is beyond the scope of the present article and it is left to a
future publication~\cite{bgqtz}. In the next section we will restrict
only to the static solutions, $r_\mp<r_\pm<r$ or $r<r_\mp<r_\pm$ for
$k=1$ and $0<r<r_+$ for $k=0$ and $k=-1$, keeping in mind their
possible relevance for cosmology as well as the self-tuning mechanism
for the cosmological constant and the possible bulk violations of the
4D Lorentz invariance.

Before finishing this section a comment is in order. In the general
action above we have not included a cosmological constant term in the
bulk.  The reason is the following. The term would be of the form $V
e^{\nu\phi}$ with $V$ some constant. However it has been
shown~\cite{wiltshire} that for $\nu\neq 0$, and for $\nu=0$ but $V >
0$, there are no solutions of the field equations consistent with the
symmetries we imposed. The only possibility would be to have $\nu=0$
and $V<0$ for which some numerical solutions are known, or to restrict
to the case of no dilaton couplings $b=0$ which would reduce to the
case with only gauge fields and no dilatons already considered in the
literature. There still remains the possibility of a nontrivial
dilaton potential $V(\phi)$ with a stationary point for which there
should be solutions~\cite{wiltshire}. This possibility will not be
considered here. However, it should be mentioned that, in absence of
gauge fields but with a Liouville potential for the scalar field, a
dilatonic domain wall ($k=0$) solution has been constructed by Cai and
Zhang in~\cite{CZ}.

\section{Codimension one brane worlds}\label{sec3}

In order to incorporate the external $n$-brane in the bulk geometry
described before we will restrict to the case of a codimension one
brane world which corresponds to the case $q=0$, that is a point-like
singularity.  Therefore we are restricted to the case of an $n$-brane
in $d=n+2$ spacetime dimensions.

To be concrete, let us define the starting action for this model. It
consists of the sum of bulk action as before (setting $q=0$ and fixing
the values of $\alpha$, $\lambda$ and $\eta$) plus a brane part.  We
have, in the Einstein frame:
\begin{equation}
\label{actionn}
S = \frac{1}{2 \kappa_{n+2}^2} \int {d^{n+2}x \, \sqrt{g_{n+2}} \,
\left( R -\frac{1}{2}(\nabla \phi)^{2} - 2\kappa_{n+2}^2 e^{-\sigma
\phi} F_{\mu\nu}F^{\mu\nu} \right)} + S_{\rm br} + S_{\rm G.H.}  \,,
\end{equation}
with
\begin{equation}
S_{\rm br} = - \int {d^{n+1}x \,\sqrt{g_{n+1}\, f^{n+1}(\phi)} \,
{\mathcal{L}}_m (\psi, \nabla\psi, g_{ab} f(\phi))}\, z(\phi) \, ,
\end{equation}
as before, $\phi$ is the dilaton field, while $F_{\mu\nu}=\partial_\mu
A_\nu-\partial_\nu A_\mu$ is the field strength tensor of a gauge
field, $A_\mu$; $S_{G.H.}$ is the Gibbons-Hawking term and ${\mathcal
{L}}_m$ is the lagrangian for the matter fields, $\psi$, living on the
brane which we model as a perfect fluid; $\kappa_{n+2}^2$ is the
effective Newton constant in $(n+2)$ dimensions.  The matter on the
brane couples to the bulk through gravity or {via} the dilaton
only. The conformal coupling of the dilaton to matter is specified by
the function $f(\phi)$ and we have also introduced a multiplicative
coupling through the function $z(\phi)$.  Varying the action
(\ref{actionn}) we will obtain an extra term coming from the brane
source in the equations of motion for the metric and the dilaton. They
are as follows ($r=\mathcal{R}(\tau)$ is the position of the brane):
\begin{eqnarray}
G_{\mu\nu} &=& -\frac{1}{4}(\nabla\phi)^2\, g_{\mu\nu} +\frac{1}{2}\,
\nabla_{\mu}\phi\, \nabla_{\nu}\phi
\nonumber \\&&
+\,2\kappa_{n+2}^2\,e^{-\sigma\phi} \left( 2 F_\mu{}^\lambda
F_{\nu\lambda} - \frac{1}{2} g_{\mu\nu}F^2 \right) +\kappa_{n+2}^2
\sqrt{\frac{g_{n+1}}{g_{n+2}}} T_{ab} \delta^a_\mu\delta_\nu^b
\delta(r-{\mathcal R}(\tau))\,,
\label{eq:BraneEinstein}\\
\nabla^2 \phi & = & -2\kappa_{n+2}^2\,\sigma e^{-\sigma\phi} F^2 -
2\kappa_{n+2}^2  \sqrt{\frac{g_{n+1}}{g_{n+2}}}
\left((np-\rho)\frac{f'}{2f}+\omega_{\mathcal{L}}\rho \frac{z'}{z}
\right)  \delta (r-{\mathcal R}(\tau))\,.\qquad
\label{eq:BraneDilaton}
\end{eqnarray}
$T_{ab}$ is the brane stress-energy tensor coupled to the induced
metric on the brane and it is given by
\begin{equation}
\label{energytensor4d}
T^{a}{}_{b} = \frac{2}{\sqrt{g_{n+1}}}\, \frac{\delta S_{\rm
br}}{\delta g_{ac}} \, g_{cb} = {\rm diag} (-\rho, p,\ldots, p) =
f^{(n+1)/2}(\phi) z(\phi)\, {\rm diag} (-\bar{\rho}, \bar{p},\ldots,
\bar{p}) \,,
\end{equation}
where $\bar{\rho}$ and $\bar{p}$ would be the energy density and
pressure of the brane in absence of coupling to the dilaton, while the
physical energy density and pressure coupled to the induced metric in
the Einstein frame are $\rho$ and $p$. The action for the matter on
the brane is expressed in terms of the energy density through the
parameter $\omega_{\mathcal{L}}$:\footnote{The value of
$\omega_{\mathcal{L}}$ depends on the type of matter of the
brane. Known cases correspond to $\omega_{\mathcal{L}}=-\omega$ for a
time dependent scalar field \cite{kolb}.  There are also other cases
for which $\omega_{\mathcal{L}}=1$ \cite{H-E}.}
\begin{equation}
f^{(n+1)/2}(\phi) \, z(\phi) \, {\mathcal{L}}_m (\psi, \nabla\psi,
g_{ab} f(\phi)) = -\omega_{\mathcal{L}}\, \rho\, .
\end{equation}
Notice that the energy-momentum conservation reads
\begin{equation}
\nabla_\mu T^{\mu\nu} =0\,; 
\end{equation}
where $T^{\mu\nu}$ is the total $n+2$-dimensional stress-energy tensor.

Ignoring for the moment the presence of the brane, the solution of the
bulk corresponds exactly to the $q=0$ case of the first section. This
is given by
\begin{eqnarray}
ds^2&=& -h_+ h_-^{1-(n-1)b}dt^2 +h_+^{-1} h_-^{-1+b} dr^2 +r^2 h_-^{b}
dx^2_{n,k} \,,
\\
\phi&=&\frac{(n-1) b}{\sigma} \ln h_-\,,
\\
F_{tr}&=&\frac{Q}{r^{n}}\,,
\label{eq:metric(n+2)D}
\end{eqnarray}
where now the expressions for $b$ and $Q$ read
\begin{eqnarray}
b &=& \frac{2 n \sigma^2}{(n-1)(n\sigma^2+ 2(n-1))} \,,
\label{eq:bsigma}\\
Q^2 &=& \frac{n(n-1)^2 (r_+r_-)^{n-1}}{2\kappa_{n+2}^2( n \sigma^2 +
2(n-1))} \,.
\end{eqnarray}
And the functions $h_\pm$ are the sign amended harmonic functions
(\ref{eq:h+-def2}).  Note that depending on the value of the dilaton
to the gauge field, the parameter $b$ varies between $0$ and
$2/(n-1)$. Thus, even in the cases $k=0,-1$ where $h_-$ is decreasing
with $r$, the spatial warp factor $r^2h_-^b$ always remains
increasing: there is no bounce.

\subsection{Introducing the brane}

Let us now introduce a dynamical $n$-brane moving into the static
$n+2$ bulk described in the last section. The brane will be separating
two regions of the above discussed background; that is, we are gluing
two slices of the metric together at the position of the brane. We
will assume that our solutions possess a $\mathbb{Z}_2$ symmetry
``centered'' at the brane (this means essentially we change
$r\rightarrow \mathcal{R}^2/r$ on the two sides of the brane and
therefore identify the two sides of the spacetime in that dimension,
for a detailed discussion see~\cite{CEG,grojean}).
 
In order to do this, we have to satisfy the Israel junction conditions
at the brane as follows~\cite{israel}.  Consider a general boundary
$X^\mu$, parametrized by the cosmic time~$\tau$,
\begin{equation}
X^\mu=(t(\tau), {\mathcal R}(\tau), x_1, \dots, x_n)\,,
\end{equation}
such that the induced metric on the boundary becomes
\begin{equation}
\label{branemetricn}
ds^2 = -d\tau^2 + a^2(\tau) dx_{n,k}^2\,,
\end{equation}
with $a(\tau)$ is the scale factor on the brane, i.e.
\begin{equation}
a(\tau) =  r\, h_-^{b/2}(r)_{\bigl|r={\mathcal
R}(\tau)}
\end{equation}
and the cosmic time $\tau$ is defined such that (a dot means
derivative with respect to the proper time)
\begin{equation}
\label{tpunton}
- h_+ h_-^{1-(n-1)b} \dot t^2  +  h_+^{-1} h_-^{b-1}\dot {\mathcal
R}^2 = -1\,.
\end{equation}
The components of the extrinsic curvature have to satisfy
the junction conditions:
\begin{equation}
\left[ K_{ab} \right]_-^+ =- \kappa^2_{n+2}\left( T_{ab} - \frac{1}{n}
g_{ab} T^c{}_c \right),
\end{equation}
while the junction condition for the dilaton derives from
(\ref{eq:BraneDilaton}) and reads ($u$ is the unit normal to the
brane)~\cite{ScalarGravity}:
\begin{equation}
\left[ u.\partial \phi\right]_-^+ = 2 \kappa^2_{n+2}
 \left(-(np-\rho)\frac{f'}{2f}-\omega_{\mathcal{L}}\,\rho\,
 \frac{z'}{z} \right).
\end{equation}
Notice that for the gauge field there are no junction conditions since
it couples only to the bulk and not to the brane.  With a
$\mathbb{Z}_2$ symmetry between the two sides of the brane the
junction equations are:
\begin{eqnarray}
\rho &=& \mp 2n\kappa^{-2}_{n+2} \left( \frac{1}{{\mathcal R}} +
\frac{b\,h_-'}{2\,h_-} \right) h_+^{1/2} h_-^{(1-b)/2}
\sqrt{1+h_+^{-1} h_-^{-1+b} {\dot{{\mathcal R}}}^2} \,,\qquad~
\label{eq:Jump1}\\
n p + (n-1) \rho &=&\pm 2n\kappa^{-2}_{n+2} \frac{h_+^{-1/2}
h_-^{(b-1)/2} }{\sqrt{1+h_+^{-1} h_-^{-1+b} {\dot{{\mathcal R}}}^2}}
\Biggl( {\ddot{{\mathcal R}} - \frac{(n-2)b h_-'}{2\,h_-}
{\dot{{\mathcal R}}}^2}
\nonumber\\&&
\qquad+\frac{1}{2} h_-^{1-b} h_+' + \left(\frac{1}{2} - \frac{n-1}{2}b\right) h_+
h_-^{-b} h_-' \Biggr),
\label{eq:Jump2}\\
(np-\rho)\frac{f'}{2f} +\omega_{\mathcal{L}}\rho\frac{z'}{z} &=&
\mp (n-1)\kappa^{-2}_{n+2} \frac{b}{\sigma} h_-' h_+^{1/2}
h_-^{-(1+b)/2} \sqrt{1+h_+^{-1} h_-^{-1+b} {\dot{{\mathcal R}}}^2}\,.
\label{eq:Jump3}
\end{eqnarray}
Note that due to the absence of bounce in the spatial warp factor $r^2
h_-^b$, there is a simple connection between the sign of $\rho$ and
the region of the spacetime cut by the $\mathbb{Z}_2$ symmetry: for
positive (negative) energy density (lower (upper) signs in the
previous eqs.) we keep the interior (exterior) region of the
background, $r<{ \mathcal R}$ ($r>{ \mathcal R}$).

Even if quite messy, these junction equations have nice physical
interpretation. Indeed a first combination of
(\ref{eq:Jump1})--(\ref{eq:Jump3}) gives us the (non)conservation
equation for the energy on the brane,
\begin{equation}
\label{eq:conservation}
\dot{\rho} + n (\rho + p) H = -\left( (np-\rho)\frac{f'}{2f} +
\omega_{\mathcal{L}}\,\rho\, \frac{z'}{z} \right)\dot{\phi} \,,
\end{equation}
where $H$ is the Hubble parameter, i.e., the time variation of $a={
\mathcal R}\, h_-^{b/2}$, the scale factor on the brane:
\begin{equation}
\label{hubble}
H(\tau) = \frac{\dot a(\tau)}{a(\tau)} = \left( \frac{1}{{ \mathcal
R}}+\frac{b\,h_-'}{2\,h_-} \right)\, \dot{{ \mathcal R}} \,.
\end{equation}
Another combination gives a Friedmann-type equation that relates the
Hubble parameter to the energy density on the brane:
\begin{equation}
\label{eq:Friedmann}
H^2 = \frac{\kappa_{n+2}^4}{4 n^2}\,\rho^2 - \left( \frac{1}{a(\tau)}+
\frac{b\,h_-'}{2\,h_-^{1+b/2}} \right)^2 h_+h_- .
\end{equation}
Finally a third combination will characterize the equation of state of
the matter on the brane: $p=\omega \rho$ with
\begin{equation}
\label{eq:EqState}
\omega = \left( \frac{(n-1)\, b\, h_-'}{2n\, \sigma\, h_-} \left(
\frac{1}{{\mathcal R}} + \frac{b\, h_-'}{2\, h_-} \right)^{-1}
-\omega_{\mathcal{L}} \frac{z'}{z} +\frac{f'}{2\,f} \right)
\frac{2\,f}{n\,f'} \,.
\end{equation}
In general the position of the brane will be time-dependent and so
also the equation of state will be. In the interesting case of
exponential couplings to the dilaton, $f=f_0\, \exp(\beta \phi)$ and
$z=z_0\, \exp(\gamma \phi)$, clearly one could obtain a constant
$\omega$ by tuning the parameters of the action such that $b=0$ or
either by choosing a constant of integration $r_-=0$.  Surprisingly
when the curvature vanishes, $k=0$, the previous formula becomes:
\begin{equation}
\label{eq:JumpW}
\omega = -\frac{2(n-1)^2\, b}{n^2\, \beta\sigma\,(2-(n-1)b)}
+\left(1-\frac{2\omega_{\mathcal{L}} \gamma}{\beta}\right) \frac{1}{n}\,,
\end{equation}
describing an equation of state that remains constant.  Since $b$ is
related through (\ref{eq:bsigma}) to the parameters that enter in the
action, this expression actually fixes completely the equation of
state from the original parameters of the lagrangian, and once
specifying those parameters there is only one possible equation of
state allowed. We may also work backwards and find which couplings in
the action allow a particularly interesting equation of state (say
$\omega=-1$).  This is the same fine-tuning of the dilaton couplings
as in the original scalar self-tuning models~\cite{ADKS,KSS}.  We will
come back to this issue in the next section.

It is easy to see that the equations (\ref{eq:conservation}),
(\ref{hubble}) and (\ref{eq:Friedmann}) reduce to the well known
expressions for the case of no dilaton coupling ($b=0$) and/or no
gauge field coupling ($r_+=0$).

Before describing in detail the geometry of the brane-bulk system, let
us rewrite the (non)con\-ser\-va\-tion eq.~(\ref{eq:conservation}) in
a more usual form at least in absence of multiplicative coupling to
the dilaton ($z=1$).  Indeed this equation takes a more simpler form
in the Jordan frame defined by a Weyl rescaling of the metric with the
conformal coupling to the dilaton:
\begin{equation}
d \bar{s}^2 = f(\phi)\, ds^2 \,,\quad \hbox{i.e.,} \quad d\bar{\tau}^2
= f(\phi)\, d\tau^2 \quad \hbox{and} \quad \bar{a}^2 = f(\phi)\, a^2\,.
\end{equation}
The energy density and pressure coupled to this metric are precisely
the quantities $\bar{\rho}$ and $\bar{p}$. Then it is easy to prove
that the (non)conservation equation reads:
\begin{equation}
\frac{d}{d\bar{\tau}} \bar{\rho} + n(\bar{\rho}+\bar{p}) \bar{H} = 0
\qquad \hbox{with}\qquad \bar{H} = \frac{1}{\bar{a}} \,
\frac{d\bar{a}}{d\bar{\tau}}\,.
\end{equation}
In the presence of a multiplicative coupling, $z\neq 1$, the right
hand side would not vanish, but as soon as $z=1$, our model is simply
a Brans-Dicke theory with localized matter on a brane and it is well
known that the usual conservation equation holds in the Jordan frame.

\subsection{Geometry of the brane-bulk system}

Now we have to decide which part of the space-time to keep due to the
$\mathbb{Z}_2$ symmetry we are requiring in order to guarantee a
compact extra dimension and then finite four-dimensional Planck scale.
This will be defined by the normal vector that we choose in the
calculation of the extrinsic curvature and it is reflected in the sign
of the energy density, as we have already explained from the junction
conditions. If the normal vector points inwards, the energy density is
positive whereas for outwards normal vector it is negative.  So if
${\mathcal R}(\tau)$ represents the location of the brane, we can glue
two interior, $r<{\mathcal R}$ (exterior, $r>{\mathcal R}$), regions
by taking a positive (negative) energy density.

\pagebreak[3]

There will be three (two) zones (see figures~\ref{fig1}
and~\ref{fig2}), defined by the black hole/0-singularity geometry,
where the brane can be located for $k=1$ ($k=0,-1$), in either case,
we take two exterior or interior regions. Let us analyze this point in
detail.
\begin{itemize}
\item \textbf{\textit{k=1} case}

\noindent (I) $r_\mp < r_\pm < {\mathcal R(\tau)}$

  {}\hskip.5cm a) $\rho<0$ Glue two exterior regions with no horizon or
    singularities in it. So extra space dimension is infinite.

  {}\hskip.5cm b) $\rho>0$ Glue two interior regions which contain
    the two singularities $r=0$, $r_-$, protected by the
    horizon  at $r_+$ if $r_+>r_-$ in which case one can naturally
    cut the space at the horizon and avoid in a  natural way the
    singularities with a positive energy density on the brane.
    However, if $r_+<r_-$ from the brane point of view, one will see
    the singularity $r_-$ and not the horizon hidden behind it.

\bigskip

\noindent (II) $r_- < {\mathcal R(\tau)} < r_+$

  {}\hskip.5cm a) $\rho<0$ Glue two exterior regions which include the
    horizon at $r_+$ and no singularities at all.

  {}\hskip.5cm b) $\rho>0$ Glue two interior regions which contain
    the two naked singularities $r=0$, $r_-$.

\bigskip

\noindent (II') $r_+ < {\mathcal R(\tau)} < r_-$

  {}\hskip.5cm a) $\rho<0$ Glue two exterior regions which include the
    the naked singularity at $r_-$.

  {}\hskip.5cm b) $\rho>0$ Glue two interior regions which contain
    the horizon at $r_+$.

\bigskip

Notice that in both cases (II) and (II') $r$ is a timelike coordinate
in the interval $[r_\mp,r_\pm]$ and so the space-time there is
cosmological. The singularity at $r_-$ is a null-singularity.

\bigskip

\noindent (III) ${\mathcal R(\tau)} < r_\mp < r_\pm$
 
        {}\hskip.5cm a) $\rho<0$ The metric becomes again static and
            one can glue two exterior regions with the singularity at
            $r_-$ and the horizon at $r_+$ that shields the
            singularity as long as $r_+<r_-$. However, if $r_-<r_+$
            then from the point of view of the brane, one will see a
            naked singularity and not the horizon, located behind the
            singularity! In both situations, one can then cut the
            space at the singularity or the horizon to get a finite
            extra dimension.

{}\hskip.5cm b) $\rho>0$ Glue two interior regions which contain the
    naked singularity at $r=0$. One can cut the space at the
    singularity there.

\bigskip

In some cases we can see that it is possible to cut the space in such
a way that naked singularities are avoided, thus getting rid of the
problems described for example in~\cite{ADKS,FLLN}. For instance we
can choose naturally the brane as described in (I)b: we then require
$r_-<r_+<{\mathcal R(\tau)}$ and a positive energy density.

\bigskip

\item \textbf{\textit{k=0} and \textit{k=--1} cases}

\noindent (I) $r_+ < {\mathcal R(\tau)}$

 {}\hskip.5cm a) $\rho<0$ The metric in this region has become
    time-dependent and is therefore cosmological.
            One can still glue two exterior regions with no
    horizon or singularities in it.

 {}\hskip.5cm b) $\rho>0$ Glue two interior regions which contain
    the singularity at $r=0$  protected by the
     horizon  at $r_+$. The bulk metric is again cosmological.

\bigskip

\noindent (II) ${\mathcal R(\tau)} < r_+$

{}\hskip.5cm a) $\rho<0$ The metric becomes again static and one can
            glue two  exterior regions keeping  the  horizon
    at  $r_+$ and then cutting there the space to obtain a
            finite extra dimension.

{}\hskip.5cm b) $\rho>0$ Glue two interior static regions which contain
    the naked singularity at  $r=0$. One can cut the space at
    the singularity getting a finite extra dimension.

\vskip 0.3 cm
\noindent
There are two possibilities to obtain a finite extradimension
and thus a finite Planck scale on the brane: putting the brane
at  ${\mathcal R}\le r_+$ and, taking a negative energy density and avoiding
the singularity at $r=0$ as described in (II)a;  or  taking a
positive energy density but  dealing again with a naked singularity
as described in (II)b. We will see later, that any of these choices
is perfectly consistent with the positivity theorems of the energy
density on our brane.

\end{itemize}

\section{The 5 dimensional model}\label{sec4}

In this section we would like to describe some interesting features of
the model presented in the case when we have a 3-brane representing
our world. In order to do that, we just have to take $n=3$ in all the
relations of the last section.

Besides simplicity and the obvious relevance of 3-branes, this will
allow us to make contact with the solutions discussed in the
literature for 5 dimensions. We will consider the various limits of
our solution to reobtain the different systems already considered in
the literature for $5$ dimensions where only gauge fields or only
dilaton fields were considered.
\begin{itemize}
\item[(i)] In the limits $r_+=b=0$ or $r_-=b=0$ both the gauge and
dilaton fields vanish and we end up with the solution given
in~\cite{BCG} with vanishing bulk cosmological constant which, for
$k=1$ corresponds to the usual Schwarzschild black hole solution.
\item[(ii)] In the limit $b=0$, one obtains a solution with a
vanishing dilaton field, that corresponds to the model found
in~\cite{CEG} again without bulk cosmological constant.
\item[(iii)] The case $r_+=0$, $b=2/3$, $k=0$ corresponds to the usual
scalar self-tuning geometry, as we will see below.
\end{itemize}

\subsection{Static 3-brane universe}

We will discuss now in detail the implications of the junction
conditions for matter on the 3-brane, determined by the energy density
and the pressure. First we will examine under which conditions the
brane remains static in the bulk: ${\mathcal R(\tau)}=R_0$.  The
static brane solution is of particular relevance in the case of a
vanishing curvature, $k=0$, since the induced metric on the brane is
then 4D Poincar\'e invariant.

We will restrict to exponential couplings between the brane and the dilaton:
\begin{equation}
f(\phi) = f_0\, e^{\beta\phi}\,, \qquad z(\phi) = z_0\, e^{\gamma\phi}\,.
\end{equation}
Thus according to the junction equation (\ref{eq:JumpW}), the equation
of state for a static brane corresponds to $\omega=$const.

Remember that for $k=0$, the only region where the brane can remain
static is $r<r_+$. In order to avoid the singularity problem we will
deal with negative energy density and glue two exterior regions
($r>R_0$).

For a static brane, the junction equations
(\ref{eq:Jump1})--(\ref{eq:Jump3}) simplify ($\dot{\mathcal
R}=\ddot{\mathcal R}=0$), leaving us with
\begin{eqnarray}
\label{staticjunctsk0}
\frac{1}{6}\,\kappa_5^{2}\,\rho &=& -\frac{(1-b)}{R_{0}}
\left(\frac{r_-}{R_0} \right)^{1-b} \sqrt{\left( \frac{r_+}{R_0}
\right)^2-1 } \,,
\\
\frac{1}{6}\,\kappa_5^{2}\, (2+3\omega) \rho & = & \frac{1}{R_0}
\left( 2b-1 - \frac{r_+^2}{r_+^2-R_0^2} \right) \left(\frac{r_-}{R_0}
\right)^{1-b} \sqrt{\left( \frac{r_+}{R_0} \right)^2-1 } \,,\qquad
\\
\label{eq:3braneW}
\omega & = & \frac{1}{3} - \frac{2\omega_\mathcal{L}\, \gamma}{3\beta}
-\frac{4\, b}{9\beta\, (1-b)\sigma} \,.
\end{eqnarray}
At this point, it should be noticed that the interpretation of one
combination of the jump equations as the conservation equation is
erroneous for a static brane and even if the conservation equation is
trivially satisfied, we really have three independent junction
conditions.  {A priori} we have also three constants of integration
$R_0,r_+$ and $r_-$.  We should mention that, the combination
(\ref{eq:3braneW}) of the jump equations involves only parameters that
appears in the action and therefore requires a kind of fine-tuning; we
will discuss this issue in the next subsection.  The two other
equations fix the two constants of integration $r_+$ and $r_{-}$:
\begin{equation}
r_+=R_0 \sqrt{1+\frac{1}{3(1-b)\omega}}\,, \qquad r_-=R_0 \left(
-\frac{1}{6} \kappa_5^2\, \rho R_0 \sqrt{3\omega\over(1-b)}
\right)^{1/(1-b)}.
\end{equation}
And the consistency of the solution that requires $r_+>R_0$ translates into
\begin{equation}
0\leq \omega\,.
\end{equation}
Notice that since the energy density on the brane is negative, the
weak energy conditions are violated anyway; in particular even if
matter on the brane can have a non exotic equation of state
($0<\omega<1$), still $p+\rho<0$.\footnote{We thank J.~Cline and
H.~Firouzjahi for discussion on this point.}

Even if of less physical relevance, a similar study can be conducted
for non vanishing curvature, $k=\pm 1$.  Let us consider, for
instance, $k=+1$. The brane, with positive energy density, will be
located in the region $r_\mp < r_\pm < R_0$. It is possible to see
that in this case it is not necessary to require fine-tuning between
the parameters of the action, and a possible expression for $\omega$
is the following:
\begin{equation}
\label{omegak1}
\omega = -\frac{1}{3} \frac{R_0^2-r_-^2}{R_0^2-(1-b)r_-^2}
\left(2+\frac{r_{-}^{2}}{R_{0}^{2}-r_{-}^{2}}+
\frac{r_{+}^{2}}{R_{0}^{2}-r_{+}^{2}}\right).
\end{equation}
The formula (\ref{omegak1}) shows that $\omega$ is always less than
zero; the physical requirement $\omega \ge -1$, moreover, is satisfied
when the following condition holds
\begin{equation} 
\label{omegamu}
(1-3b)\frac{r_{-}^{2}}{R_{0}^{2}-r_{-}^{2}} \le
 1-\frac{r_{+}^{2}}{R_{0}^{2}-r_{+}^{2}}\,.
\end{equation}
This last expression will be interesting later on.  Notice that as
long as (\ref{omegamu}) is satisfied, the weak energy conditions hold.

\subsection{Self-tuning solutions}\label{sec:selftuning}

Let us discuss the issue of the self-tuning of the cosmological
constant that has attracted some interest recently~\cite{ADKS,KSS}.
The main idea behind this proposal is that the field equations in
5-dimensions with a dilaton field allow for a solution which is 4D
Poincar\'e invariant whatever the vacuum energy on the brane is.
Furthermore there are not other maximally symmetric solutions and the
brane is such that any corrections to the cosmological constant coming
from matter loops on the brane can be absorbed into a shift of the
dilaton field, allowing the `self-tuning' of the cosmological
constant. The main problem of this proposal~\cite{ADKS,FLLN} is the
existence of naked singularities in the extra dimensions which cannot
be avoided~\cite{CEGH}.

A similar proposal has been made regarding the inclusion of gauge
fields instead of the dilaton~\cite{CEG}. Again the cosmological
constant can be self-tuned by adjusting the values of the charge and
mass of the corresponding $AdS$ black hole solution. However the
geometry of the brane-bulk system is such that the singularity of the
black hole can be shielded by a horizon only for a brane with an
exotic equation of state $\omega<-1$. In this subsection we would like
to study the possibility of obtaining a self-tuning brane in our
dilatonic backgrounds.

First we have to address an important issue regarding equation
(\ref{eq:JumpW}).  When the gauge fields are nonvanishing, we proved
that $\omega$ was completely determined by the parameters of the
lagrangian, because the constant $b$ was determined by $\sigma$, the
coupling of the dilaton to the gauge field, as in
equation~(\ref{eq:bsigma}).  This means that in order to get a
specific equation of state we would need to tune the parameters of the
action ($\gamma, \beta, \sigma$).  When the gauge fields vanish i.e.,
$Q=0$, $b$ is an arbitrary constant of integration (only limited by
$0\leq b \leq 1$) independent of the parameters of the action,
however, since $r_+=0$, the junction conditions completely determine
$b$ in terms of $\omega$ and once again (\ref{eq:JumpW}) requires a
fine-tuning between the equation of state of the matter on the brane
and the dilaton couplings. This is the same fine-tuning problem as in
the original scalar self-tuning models~\cite{ADKS,KSS}.

Actually the uncharged dilatonic background solution is given by
\begin{eqnarray}
ds^2&=& -h_+ h_-^{1-2b}dt^2 +h_+^{-1} h_-^{-1+b} dr^2 +r^2 h_-^{b}
dx^2_{3,k} \,,
\nonumber\\
\phi&=&\pm \sqrt{3b(1-b)}\, \ln h_- \,,
\label{eq:b=0metric}
\end{eqnarray}
where the functions $h_\pm$ are now given by
\begin{equation}
h_+= \mathop{\rm sgn} h\,, \qquad h_-= \left|h \right |, \qquad h=k-
\frac{l}{r^2}\,, \qquad k=0,\pm1\,.
\end{equation}
and $l$ is an integration constant which can be positive or negative
since there is no reality constraint for the electric charge anymore.
The jump equations on a brane with positive energy density, for $k=0$,
now read:
\begin{eqnarray}
\kappa_5^2 \rho &=& 6(1-b) \frac{1}{R_0} \left(\frac{-l}{R_0}\right)^{1-b}\,,
\nonumber\\
b&=&  \frac{1+3\omega}{3\omega}\,,
\nonumber\\
(3\omega-1) \frac{f'}{2f} + \omega_{\mathcal{L}} \frac{z'}{z} &=& \mp
\sqrt{\frac{b}{3(1-b)}} \,.
\end{eqnarray}

A sufficient condition for $r$ to be spacelike is the choice $k=0$ and
$l<0$; for simplicity we will take $l=-1$. Then we can realize that
the full metric (\ref{eq:b=0metric}) will be 4D Poincar\'e invariant
under the condition that the spatial and temporal warp factors have
the same dependence on $r$:
\begin{equation}
r^{-2(1-2b)}=r^2r^{-2b}\,,
\quad \hbox{i.e.,}
\quad b=\frac23\,.
\end{equation}
So, we are left with the bulk metric:
\begin{equation}
ds^2 = \sqrt{1-\left|\frac{y}{y_0}\right|}
\left(-dt^2+dx_{3,k=0}^2\right) + dy^2 \,,
\end{equation}
where the new coordinate $y$ is related to $r$ by: $r^{1/3}dr=dy$.
This is just the bulk metric of the original scalar self-tuning
model~\cite{ADKS,KSS}, corresponding to the solution II
of~\cite{KSS}. It can easily be verified that for $\beta = 1/\sqrt{6}$
and $\gamma=0$ we get $\omega=-1$ which is the equation of state used
in~\cite{KSS}.\footnote{To make the comparison with~\cite{ADKS,KSS}
equations we have to redefine the dilaton field appropriately.}

We can actually go beyond the results in~\cite{ADKS,KSS} by noting
that the other solutions with different values of $b$ will also allow
for a 3-brane with a Poincar\'e invariant induced metric whatever the
value of its vacuum energy is. These other solutions will break the 4D
Poincar\'e invariance in the bulk, which is still safe for gauge
interactions of the Standard Model living on the brane, and may lead
to interesting observational consequences in gravitational waves
experiments as we will discuss in the next section.  Nevertheless
there is no way to overcome the problem of the naked singularity in
these cases since none of the self-tuning solutions with vanishing
gauge field involves a horizon.

In the case of non vanishing gauge fields, even if the value of
$\omega$ is again completely determined by the parameters of the
lagrangian ($\beta, \gamma, \sigma $), the global geometry is such
that in the case of negative energy density on the brane, there are no
naked singularities present.  Also $\omega$ can be in the physically
allowed range, which is an improvement on the cases with only dilaton
or only gauge fields.

It is easy to verify that the other maximally symmetric
four-dimensional spaces, i.e.\ the de Sitter and anti-de Sitter
geometries, cannot be obtained from our solutions.

\subsection{Violation of 4D lorentz invariance}

It has recently been emphasized~\cite{CEG,cXdim} that brane world
models can have an interesting effect regarding the speed of
propagation of light and gravitational signals. Indeed the general
solution (\ref{eq:metric}) breaks the 4D Poincar\'e invariance in the
bulk and as a consequence the speed of propagation of electromagnetic
signals parallel to the brane depends on the location of the brane. It
is then {a priori} possible to foresee signals that travel faster
through the bulk than on the brane. It has been argued for a long
time~\cite{VaryingC} that faster than light propagation and/or
variation of the speed of light can solve many of the cosmological
puzzles (horizon problem, cosmological constant problem, {etc}).
These ideas witness today a renewed interest in the context of brane
world models~\cite{cXdim}.

The violation of 4D Lorentz invariance was extensively discussed
in~\cite{CEG}, where the authors shown the correctness of the
intuitive idea for which if one has a decreasing speed of
gravitational waves moving away from the brane, then the brane Lorentz
invariance can be recovered, in the sense that the gravitational waves
prefer to move on the brane, due to the Fermat's principle.

We can now calculate in which cases we can have a negative derivative
for the speed of gravitational waves, i.e., a decreasing speed of
light away of the brane such that according to Fermat's theorem the
gravitational waves will actually propagate on the brane rather than
through the bulk.

Let us first examine the case of a static brane with a 4D Poincar\'e
invariant induced metric, which requires a vanishing curvature
$k=0$. From the expression (\ref{eq:metric(n+2)D}) of the metric, we
deduce the local speed of propagation of gravitational waves in a
direction parallel to the brane:
\begin{equation}
\label{speed0}
c^2_{\rm grav}(r)= \left(\frac{r_{+}^{2}}{r^{2}}-1\right)
\left(\frac{1}{r^{2}}\right)^{(2-3b)} r_{-}^{2(1-3b)} \,.
\end{equation}
This expression is of course valid only in the region where $r$ is a
spacelike coordinate. It is easy to see that, since $0<b<1$, the local
speed of propagation $c^2_{\rm grav}(r)$ is always a decreasing function
of $r$ in the region $0<r<r_+$, this means that this local speed of
propagation will be either decreasing or increasing away from the
brane in $R_{0}$ depending on the sign of the brane energy density
(see figure~\ref{fig3}):
\begin{itemize}
\item
Positive energy density. We keep the interior region ($r<R_0$) and
thus the speed of propagation is increasing away from the brane and
the gravitational waves will prefer to propagate through the bulk.
Note that in this case the naked singularity at $r=0$ is not shielded
by a horizon.
\item
Negative energy density. We keep now the exterior region
($R_0<r<r_+$), therefore in this case the gravitational waves will
prefer to travel through the brane instead of the bulk and there will
be no evidence of Lorentz violation.
\end{itemize}
We can then conclude that Lorentz violation would be manifest only in
the case with naked singularity, similar to the situation found
in~\cite{CEG}.

We can also consider the case $k=1$ (although it is less interesting
since the induced metric on the brane is not Poincar\'e invariant). In
this case the conclusion regarding the increase or decrease of the
speed of gravitational waves is different. The speed of the
gravitational waves as a function of the position in the bulk is given
by the following expression
\begin{equation}\label{vel}
c_{\rm grav}^2(r)= \left(1-\frac{r_{+}^{2}}{r^{2}}\right)
\left(1-\frac{r_{-}^{2}}{r^{2}}\right)^{1-3b}r^{-2}\, ,
\end{equation}
and we have a brane with  positive energy density located in the
region $r_+<R_0$.

The condition of decreasing speed of gravitational waves away from the
brane is satisfied when
\begin{equation}
\label{speed}
R_0^{2}+3(1-b)\frac{r_{-}^{2}r_{+}^{2}}{R_0^{2}} \le 2 r_{+}^{2}
+(2-3b)r_{-}^{2}\,.
\end{equation}
The behaviour of the speed of gravitational waves is clear from the
plots in figure~\ref{fig4}. It is very interesting to note that this
condition is exactly the same as (\ref{omegamu}), but with the
opposite sign. The conclusion is that, with $k=1$, to have a
decreasing speed of light moving away from the brane one needs to
consider an exotic form of matter living on the brane, with $w \le
-1$. {Vice-versa}, if the matter on the brane has a standard equation
of state, one must take into account possible non negligible effects
of changes in the speed of gravitational waves in the visible world.

\EPSFIGURE[t]{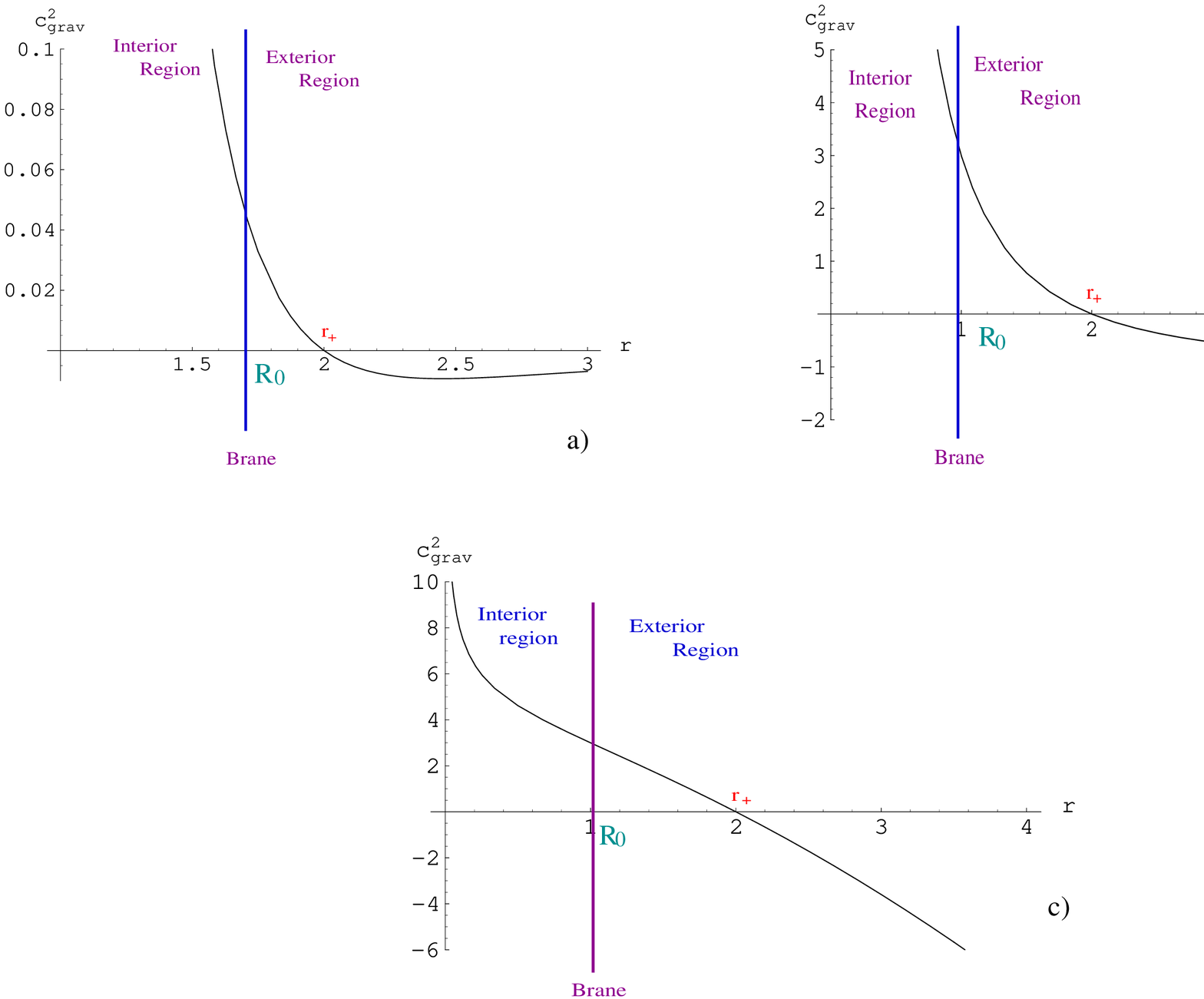, width=\textwidth}{Speed of
gravitational waves as a function of the extradimension $r$ for $k=0$
and a) $b=0.001$, b) $b=2/3$, c) $b=0.95$. The brane is always located
inside the horizon, sited at $r=r_+$. We took $r_-=1$,
$r_+=2$.\label{fig3}}

\EPSFIGURE[t]{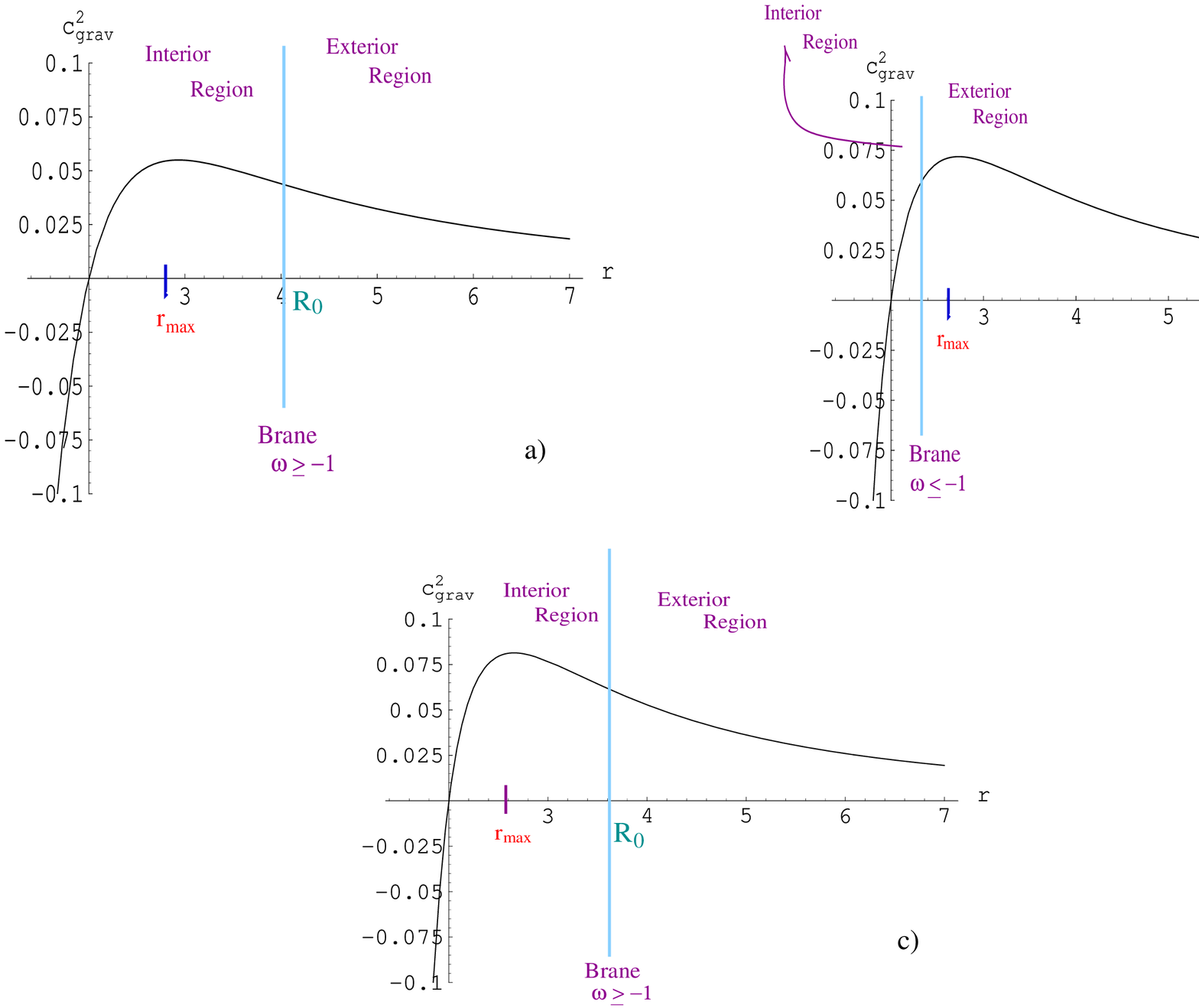, width=\textwidth}{Speed of
gravitational waves as a function of the extradimension $r$ for $k=1$
and a) $b=0.001$, b) $b=2/3$, c) $b=0.95$. Here is clear that for a
non exotic matter living on the brane we need to locate the brane to
the right hand side of the maximum, having then Lorentz violation
effects (we always keep the interior region: $r<R_0$).  We took
$r_-=1$, $r_+=2$.\label{fig4}}

\section{Conclusions and comments}\label{sec5}

We have started a general treatment of higher dimensional singular
geometries for which external branes of arbitrary dimension can be
incorporated.  There are two kinds of branes in our solutions that
should not be confused. First are the $q$ dimensional bulk
singularities that in the $k=1$ case are black branes and in general
are `$q$-brane singularities'. The second brane is an external
$n$-brane that is added given that the geometry factorizes (with a
warp factor) between the $q$ dimensional part and the $n+2$
dimensional for which the $n$~dimensional hypersurfaces are maximally
symmetric with constant curvature. We may see this, in the case $k=1$,
as a brane in a background defined by a second, orthogonal, brane. In
the extremal case ($r_-=r_+$) the bulk solution is supersymmetric and
there may be an interesting reformulation of this configuration in
terms of the AdS/CFT correspondence. For $k=0,-1$ there is no analogue
to an extremal case: there are no time-like Killing vectors and there
does not seem to be a supersymmetric limit.

\looseness=1Besides their intrinsic relevance, these configurations can be seen as
the starting point of brane cosmology in the higher dimensional space.
We have found several interesting properties of our solutions which
are novel. First the $k=0,-1$ bulk solutions are to the best of our
knowledge new\footnote{See however~\cite{BF,similar} for related
solutions.}  and have very interesting geometrical structure having a
natural cosmological interpretation but with a time-like singularity
and a Cauchy horizon.  Similar geometries have appeared in the
past. The Penrose diagram of the Taub-NUT solution has the same
signature change, as we have, in the Cauchy horizon, but in that case
there is another horizon rather than a singularity which allows the
infinite extension of the diagram~\cite{TAUB}. A similar behaviour
appears in the study of tilted Bianchi cosmologies. In these cases the
Cauchy surface corresponds actually to a non-scalar singularity
usually referred to as a `whimper' or intermediate singularity, where
the metric changes signature \cite{ellis}.  The origin of this
intermediate singularity is the fact that the fluid matter flows in a
direction which is not perpendicular to the homogeneity surfaces. In
our case there is no fluid in the bulk and the intermediate surface is
not a singularity but a horizon.  It is interesting to notice that if
an observer in the region beyond the horizon extrapolates back in time
she/he never finds a big-bang singularity but the horizon.

Even though a detailed study of the cosmology of these configurations
is beyond our scope, we would like to point out the following
interesting properties.  For $k=0$ in the 5d bulk cosmological region,
at late times ($t\gg r_+$ so that $h_+ \sim -1$) the three standard
spatial dimensions always expand while the fifth has different
behavior depending on the values of $b$, similar to the Kasner
solution: For $2/3<b<1$ all $4$ spatial dimensions expand, the 5th
faster than the rest.  For the critical case $ b=2/3$ the four spatial
dimensions expand at the same rate.  In the interval $1/2<b<2/3$ three
dimensions expand faster than the 5th, whereas for $b=1/2$ the 5th
dimension is static and the others expand.  Finally for $0<b<1/2$ the
5th dimension contracts and the others expand.  All of this in the
Einstein frame.

Second our solutions provide new examples of violation of 4D Lorentz
invariance that may have interesting consequences regarding
gravitational waves experiments and cosmology. Finally the
introduction of the $n$ brane on the bulk static regions can be done
in such a way that the singularities can be avoided. A substantial
improvement on the situations with only gauge fields or only dilaton
regarding the self tuning mechanism for the cosmological constant was
achieved.  It remains to be seen to what extent this is real progress
in the approach to the cosmological constant problem and to what
extent it is realistic to have a negative energy density brane world.

Many things remain to be explored in this subject.  The possibility of
a nontrivial dilaton potential with a stationary point is an open
question. Also the incorporation of further antisymmetric tensors of
different rank as well as the consideration of codimension larger than
one brane worlds and the addition of more than one external branes are
left to the future.  Probably a more direct extension of the present
work is the actual use of these geometries to a detailed cosmological
study of this class of brane worlds.

We have made a step forward in studying the structure of a general
class of brane models with properties close to what we natural expect
in string theory. Their interesting properties makes them worth of
further investigation, especially in the context of brane
cosmology. We hope to report on progress in this direction in the near
future.

\acknowledgments

We would like to thank J.~Barrow, G.~Gibbons, D.~Mateos and
P.~Townsend for useful discussions. The research of F.Q. is supported
by PPARC.  I.Z.C. is supported by CONACyT (Mexico), Trinity College
(Cambridge) and COT (Cambridge Overseas Trust).  C.G. is supported in
part by the Director, Office of Science, Office of High Energy and
Nuclear Physics, of the US Department of Energy under Contract
DE-AC03-76SF00098, and in part by the National Science Foundation
under grant PHY-95-14797. G.T. is partially supported by the European
TMR project ``Across the Energy Frontier" under the contract
HPRN-CT-2000-0148.

\newcommand{\mathAG}[1]{\href{http://xxx.lanl.gov/abs/math.AG/#1}{\tt
math.AG/#1}}

\renewcommand\baselinestretch{1.09}


\normalsize

\begin{thebibliography}{99}

\bibitem{early}
K.~Akama, \emph{An early proposal of `brane world'}, \emph{Lect.\ Notes
  Phys.\ }{\bf 176} (1982) 267 [\hepth{0001113}];\\
V.A. Rubakov and M.E. Shaposhnikov, \emph{Do we live inside a domain
  wall?}, \plb{125}{1983}{136};\\
M.~Visser, \emph{An exotic class of Kaluza-Klein models},
\plb{159}{1985}{22} [\hepth{9910093}];\\
G.W. Gibbons and D.L. Wiltshire, \emph{Space-time as a membrane in
  higher dimensions}, \npb{287}{1987}{717};\\
N.~Arkani-Hamed, S.~Dimopoulos and G.~Dvali, \emph{The hierarchy
  problem and new dimensions at a millimeter}, \plb{429}{1998}{263}
  [\hepph{9803315}];
\emph{Phenomenology, astrophysics and cosmology of theories with
  sub-millimeter dimensions and tev scale quantum gravity},
  \prd{59}{1999}{086004} [\hepph{9807344}];\\
I.~Antoniadis, N.~Arkani-Hamed, S.~Dimopoulos and G.~Dvali, \emph{New
  dimensions at a millimeter to a fermi and superstrings at a TeV},
  \plb{436}{1998}{257} [\hepph{9804398}];\\
K.R. Dienes, E.~Dudas and T.~Gherghetta, \emph{Extra spacetime
  dimensions and unification}, \plb{436}{1998}{55}
  [\hepph{9803466}];
\emph{Grand unification at
  intermediate mass scales through extra dimensions},
  \npb{537}{1999}{47} [\hepph{9806292}];\\
Z.~Kakushadze and S.H.H. Tye, \emph{Brane world},
\npb{548}{1999}{180} [\hepth{9809147}];\\
L.~Randall and R.~Sundrum, \emph{A large mass hierarchy from a small
extra dimension}, \prl{83}{1999}{3370} [\hepph{9905221}];
\emph{An alternative to compactification}, \prl{83}{1999}{4690}
[\hepth{9906064}].


\bibitem{branecosmology}
P.~Bin\'etruy, C.~Deffayet and D.~Langlois, \emph{Non-conventional
  cosmology from a brane-universe}, \npb{565}{2000}{269}
  [\hepth{9905012}];\\
C.~Cs\'aki, M.~Graesser, C.~Kolda and J.~Terning, \emph{Cosmology of
  one extra dimension with localized gravity}, \plb{462}{1999}{34}
  [\hepph{9906513}];\\
J.M. Cline, C.~Grojean and G.~Servant, \emph{Cosmological expansion in
  the presence of extra dimensions}, \prl{83}{1999}{4245}
  [\hepph{9906523}];\\
E.E. Flanagan, S.H.H. Tye and I.~Wasserman, \emph{A cosmology of the
  brane world}, \prd{62}{2000}{024011} [\hepph{9909373}];\\
D.~Ida, \emph{Brane-world cosmology}, \jhep{09}{2000}{014}
[\grqc{9912002}];\\
R.N. Mohapatra, A.~Perez-Lorenzana and C.A. de~Sousa~Pires,
  \emph{Cosmology of brane-bulk models in five dimensions},
  \ijmpa{16}{2001}{1431} [\hepph{0003328}];\\
J.~Lesgourgues, S.~Pastor, M.~Peloso and L.~Sorbo, \emph{Cosmology of
  the Randall-Sundrum model after dilaton stabilization},
  \plb{489}{2000}{411} [\hepph{0004086}].

\bibitem{radion}
W.D. Goldberger and M.B. Wise, \emph{Modulus stabilization with bulk
  fields}, \prl{83}{1999}{4922} [\hepph{9907447}];\\
C.~Cs\'aki, M.~Graesser, L.~Randall and J.~Terning, \emph{Cosmology of
  brane models with radion stabilization}, \prd{62}{2000}{045015}
  [\hepph{9911406}];\\
P.~Kanti, I.I. Kogan, K.A. Olive and M.~Pospelov, \emph{Single-brane
  cosmological solutions with a stable compact extra dimension},
  \prd{61}{2000}{106004} [\hepph{9912266}];\\
J.M. Cline and H.~Firouzjahi, \emph{5-dimensional warped cosmological
  solutions with radius stabilization by a bulk scalar},
  \plb{495}{2000}{271} [\hepth{0008185}];\\
B.~Grinstein, D.R. Nolte and W.~Skiba, \emph{Radion stabilization by
  brane matter}, \prd{63}{2001}{105016} [\hepth{0012202}].

\bibitem{ccXdim}
P.J. Steinhardt, \emph{General considerations of the cosmological
  constant and the stabilization of moduli in the brane-world
  picture}, \plb{462}{1999}{41} [\hepth{9907080}];\\
C.P. Burgess, R.C. Myers and F.~Quevedo, \emph{A naturally small
  cosmological constant on the brane?}, \plb{495}{2000}{384}
  [\hepth{9911164}];\\
S.P. de~Alwis, \emph{Brane world scenarios and the cosmological
  constant}, \npb{597}{2001}{263} [\hepth{0002174}];\\
J.-W. Chen, M.A. Luty and E.~Ponton, \emph{A critical cosmological
  constant from millimeter extra dimensions}, \jhep{09}{2000}{012}
  [\hepth{0003067}];\\
C.~Schmidhuber, \emph{Micrometer gravitinos and the cosmological
  constant}, \npb{585}{2000}{385} [\hepth{0005248}];\\
Z.~Kakushadze, \emph{Why the cosmological constant problem is hard},
\plb{488}{2000}{402} [\hepth{0006059}];\\
A.~Kehagias and K.~Tamvakis, \emph{A self-tuning solution of the
  cosmological constant problem}, \hepth{0011006};\\
J.E. Kim, B.~Kyae and H.M. Lee, \emph{A model for self-tuning the
  cosmological constant}, \prl{86}{2001}{4223} [\hepth{0011118}];\\
J.M. Cline and H.~Firouzjahi, \emph{A small cosmological constant from
  warped compactification with branes}, \plb{514}{2001}{205}
  [\hepph{0012090}];\\
E.~Flanagan, N.~Jones, H.~Stoica, S.H.~Tye and I.~Wasserman, \emph{A
brane world perspective on the cosmological constant and the hierarchy
problems}, \prd{64}{2001}{045007} [\hepth{0012129}];\\
W.S.~Bae, Y.M.~Cho and S.~Moon, \emph{Living inside the horizon of the
D3-brane}, \jhep{03}{2001}{039} [\hepth{0012221}];\\
Z.~Berezhiani, M.~Chaichian, A.B.~Kobakhidze and Z.H.~Yu,
\emph{Vanishing of cosmological constant and fully localized gravity
in a brane world with extra time(s)}, \hepth{0102207};\\
P.~Brax and A.C.~Davis, \emph{Cosmological evolution on self-tuned
branes and the cosmological constant}, \jhep{05}{2001}{007}
[\hepth{0104023}];\\
K.R. Dienes, \emph{Solving the hierarchy problem without supersymmetry
  or extra dimensions: an alternative approach}, \hepph{0104274};\\
S.M. Carroll and L.~Mersini, \emph{Can we live in a self-tuning universe?},
  \hepth{0105007}.

\bibitem{ADKS}
N.~Arkani-Hamed, S.~Dimopoulos, N.~Kaloper and R.~Sundrum, \emph{A
small cosmological constant from a large extra dimension},
\plb{480}{2000}{193} [\hepth{0001197}].

\bibitem{KSS}
S.~Kachru, M.~Schulz and E.~Silverstein, \emph{Self-tuning flat domain
  walls in 5d gravity and string theory}, \prd{62}{2000}{045021}
  [\hepth{0001206}];
\emph{Bounds on curved domain walls in 5d gravity},
\prd{62}{2000}{085003}
[\hepth{0002121}].


\bibitem{CEGH}
C.~Cs\'aki, J.~Erlich, C.~Grojean and T.~Hollowood, \emph{General
  properties of the self-tuning domain wall approach to the
  cosmological constant problem}, \npb{584}{2000}{359}
  [\hepth{0004133}].

\bibitem{CEG}
C.~Cs\'aki, J.~Erlich and C.~Grojean, \emph{Gravitational Lorentz
  violations and adjustment of the cosmological constant in
  asymmetrically warped spacetimes}, \npb{604}{2001}{312}
  [\hepth{0012143}];
\emph{The cosmological constant problem in brane-worlds and
  gravitational Lorentz violations}, \grqc{0105114}.

\bibitem{fromstrings}
J.~Polchinski, \emph{Dirichlet-branes and Ramond-Ramond charges},
\prl{75}{1995}{4724} [\hepth{9510017}];\\
P.~Ho\v rava and E.~Witten, \emph{Eleven-dimensional supergravity on a
  manifold with boundary}, \npb{475}{1996}{94} [\hepth{9603142}];\\
J.D. Lykken, \emph{Weak scale superstrings}, \prd{54}{1996}{3693}
[\hepth{9603133}];\\
A.~Lukas, B.A. Ovrut, K.S. Stelle and D.~Waldram, \emph{The universe
  as a domain wall}, \prd{59}{1999}{086001} [\hepth{9803235}];\\
G.~Shiu and S.H.H. Tye, \emph{TeV scale superstring and extra
  dimensions}, \prd{58}{1998}{106007} [\hepth{9805157}];\\
H.S. Reall, \emph{Open and closed cosmological solutions of Ho\v
  rava-Witten theory}, \prd{59}{1999}{103506} [\hepth{9809195}];\\
C.P. Burgess, L.E. Ib\'a\~nez and F.~Quevedo, \emph{Strings at the
  intermediate scale or is the fermi scale dual to the Planck scale?},
  \plb{447}{1999}{257} [\hepph{9810535}];\\
S.P.~de Alwis, A.T.~Flournoy and N.~Irges, \emph{Brane worlds, the
cosmological constant and string theory}, \jhep{01}{2001}{027}
[\hepth{0004125}];
\emph{Brane worlds, the cosmological constant and string theory},
  \jhep{01}{2001}{027} [\hepth{0004125}].

\bibitem{models}
G.~Aldazabal, L.E. Ib\'a\~nez and F.~Quevedo, \emph{Standard-like
  models with broken supersymmetry from type-I string vacua},
  \jhep{01}{2000}{031} [\hepth{9909172}];
\emph{A D-brane alternative to the MSSM}, \jhep{02}{2000}{015}
  [\hepph{0001083}];\\
G.~Aldazabal, L.E. Ib\'a\~nez, F.~Quevedo and A.M. Uranga,
  \emph{D-branes at singularities: a bottom-up approach to the string
  embedding of the standard model}, \jhep{08}{2000}{002}
  [\hepth{0005067}];\\
R.~Blumenhagen, L.~Goerlich, B.~Kors and D.~Lust, \emph{Noncommutative
  compactifications of type-I strings on tori with magnetic background
  flux}, \jhep{10}{2000}{006} [\hepth{0007024}];\\
G.~Aldazabal, S.~Franco, L.E. Ib\'a\~nez, R.~Rabadan and A.M. Uranga,
  \emph{$D = 4$ chiral string compactifications from intersecting
  branes}, \hepth{0011073};\\
R.~Blumenhagen, B.~K\"ors and D.~L\"ust, \emph{Type-I strings with F-
  and B-flux}, \jhep{02}{2001}{030} [\hepth{0012156}];\\
L.E. Ib\'a\~nez, F.~Marchesano and R.~Rabadan, \emph{Getting just the
  standard model at intersecting branes}, \hepth{0105155};\\
R.~Donagi, B.A. Ovrut, T.~Pantev and D.~Waldram, \emph{Standard model
  vacua in heterotic M-theory}, \hepth{0001101};\\
R.~Donagi, B.A. Ovrut, T.~Pantev and D.~Waldram, \emph{Standard models
  from heterotic M-theory}, \hepth{9912208}; \emph{Standard-model
  bundles}, \mathAG{0008010}.

\bibitem{mirjam2}
M.~Cveti\v c and J.~Wang, \emph{Vacuum domain walls in d-dimensions: local and
  global space-time structure},  \prd{61}{2000}{124020} [\hepth{9912187}];\\
see also M.~Cveti\v c and H.H. Soleng, \emph{Naked singularities in
  dilatonic domain wall space times}, \prd{51}{1995}{5768}
  [\hepth{9411170}];
\emph{Supergravity domain walls}, \prep{282}{1997}{159}
[\hepth{9604090}].

\bibitem{kraus}
P.~Kraus, \emph{Dynamics of anti-de~Sitter domain walls},
\jhep{12}{1999}{011} [\hepth{9910149}].

\bibitem{KeKi}
A.~Kehagias and E.~Kiritsis, \emph{Mirage cosmology},
\jhep{11}{1999}{022} [\hepth{9910174}].

\bibitem{BCG}
P.~Bowcock, C.~Charmousis and R.~Gregory, \emph{General brane
  cosmologies and their global spacetime structure},
  \cqg{17}{2000}{4745} [\hepth{0007177}].

\bibitem{HS}
G.T. Horowitz and A.~Strominger, \emph{Black strings and p-branes},
\npb{360}{1991}{197}.

\bibitem{GM}
G.W. Gibbons and K.-I. Maeda, \emph{Black holes and membranes in
  higher dimensional theories with dilaton fields},
  \npb{298}{1988}{741};\\
M.J. Duff, R.R. Khuri and J.X. Lu, \emph{String solitons},
\prep{259}{1995}{213} [\hepth{9412184}];\\
D.~Youm, \emph{Black holes and solitons in string theory},
\prep{316}{1999}{1} [\hepth{9710046}].

\bibitem{miriam}
R.~Kallosh, A.~Linde, T.~Ortin, A.~Peet and A.~Van~Proeyen,
  \emph{Supersymmetry as a cosmic censor}, \prd{46}{1992}{5278}
  [\hepth{9205027}];\\
M.~Cveti\v c and A.A. Tseytlin, \emph{Solitonic strings and BPS
  saturated dyonic black holes}, \prd{53}{1996}{5619}
  [\hepth{9512031}], erratum \ibid{D 55}{1996}{3907}.

\bibitem{bgqtz}
P.~Bin\'etruy, C.~Grojean, F.~Quevedo, G.~Tasinato, I.~Zavala~C., 
 in preparation.

\bibitem{BF}
K.~Behrndt and S.~F\"orste, \emph{String Kaluza-Klein cosmology},
\npb{430}{1994}{441} [\hepth{9403179}].

\bibitem{wiltshire}
S.J. Poletti, J.~Twamley and D.L. Wiltshire, \emph{Charged dilaton
  black holes with a cosmological constant}, \prd{51}{1995}{5720}
  [\hepth{9412076}];\\
D.L. Wiltshire, \emph{Dilaton black holes with a cosmological term},
  \emph{J.\  Austral.\ Math.\ Soc.\ }{\bf B41} (1999) 198
  [\grqc{9502038}].

\bibitem{CZ}
R.-G. Cai and Y.-Z. Zhang, \emph{Holography and brane cosmology in
  domain wall backgrounds}, \hepth{0105214}.

\bibitem{kolb}
E.W.~Kolb and M.S.~Turner, \emph{The early universe}, 
Addison Wesley, 1990

\bibitem{H-E}
G.F.R. Ellis and S.W. Hawking, \emph{The large structure of space-time}, 
Cambridge University Press, 1973

\bibitem{grojean}
C.~Grojean, \emph{T self-dual transverse space and gravity trapping},
\plb{479}{2000}{273} [\hepth{0002130}].

\bibitem{israel}
W.~Israel, \emph{Singular hypersurfaces and thin shells in general
relativity}, \nc{B44S10}{1966}{1} erratum \ibid{B 48}{1966}{463}.

\bibitem{ScalarGravity}
H.A. Chamblin and H.S. Reall, \emph{Dynamic dilatonic domain walls},
\npb{562}{1999}{133} [\hepth{9903225}];\\
K.-I. Maeda and D.~Wands, \emph{Dilaton-gravity on the brane},
\prd{62}{2000}{124009} [\hepth{0008188}];\\
A.~Mennim and R.A. Battye, \emph{Cosmological expansion on a dilatonic
  brane-world}, \cqg{18}{2001}{2171} [\hepth{0008192}].

\bibitem{FLLN}
D.~Youm, \emph{Bulk fields in dilatonic and self-tuning flat domain
  walls}, \npb{589}{2000}{315} [\hepth{0002147}];\\
S.~F\"orste, Z.~Lalak, S.~Lavignac and H.P. Nilles, \emph{A comment on
  self-tuning and vanishing cosmological constant in the brane world},
  \plb{481}{2000}{360} [\hepth{0002164}];
\emph{The cosmological constant problem from a brane-world
  perspective}, \jhep{09}{2000}{034} [\hepth{0006139}].

\bibitem{VaryingC}
J.W. Moffat, \emph{Quantum gravity, the origin of time and time's
arrow}, \emph{Found.\ Phys.\ }{\bf 23} (1993) 411 [\grqc{9209001}];\\
\emph{Superluminary universe: a possible solution to the initial value
  problem in cosmology}, \emph{Int.\ J.\ Mod.\ Phys.\ }{\bf D2} (1993)
  351 [\grqc{9211020}];\\
A.~Albrecht and J.~Magueijo, \emph{A time varying speed of light as a
  solution to cosmological puzzles}, \prd{59}{1999}{043516}
  [\astroph{9811018}];\\
J.D. Barrow, \emph{Cosmologies with varying light-speed},
  \astroph{9811022};\\
J.D. Barrow and J.~Magueijo, \emph{Varying-$\alpha $ theories and
  solutions to the cosmological problems}, \astroph{9811072};\\
J.W. Moffat, \emph{Varying light velocity as a solution to the
  problems in cosmology}, \astroph{9811390};\\
M.A. Clayton and J.W. Moffat, \emph{Dynamical mechanism for varying
  light velocity as a solution to cosmological problems},
  \plb{460}{1999}{263} [\astroph{9812481}];\\
J.~Magueijo, \emph{Covariant and locally Lorentz-invariant varying
  speed of light theories}, \prd{62}{2000}{103521} [\grqc{0007036}];\\
M.A. Clayton and J.W. Moffat, \emph{A scalar-tensor cosmological model
  with dynamical light velocity}, \plb{506}{2001}{177}
  [\grqc{0101126}];\\
J.W. Moffat, \emph{Acceleration of the universe, string theory and a
  varying speed of light}, \hepth{0105017}.

\bibitem{cXdim}
G.~Kalbermann and H.~Halevi, \emph{Nearness through an extra
  dimension}, \grqc{9810083};\\
E.~Kiritsis, \emph{Supergravity, D-brane probes and thermal super
  Yang-Mills: a comparison}, \jhep{10}{1999}{010} [\hepth{9906206}];\\
G.~Kalbermann, \emph{Communication through an extra dimension},
\ijmpa{15}{2000}{3197} [\grqc{9910063}];\\
D.J.~H. Chung and K.~Freese, \emph{Can geodesics in extra dimensions
  solve the cosmological horizon problem?}, \prd{62}{2000}{063513}
  [\hepph{9910235}];\\
S.H.~S. Alexander, \emph{On the varying speed of light in a
  brane-induced FRW universe}, \jhep{11}{2000}{017}
  [\hepth{9912037}];\\
H.~Ishihara, \emph{Causality of the brane universe},
\prl{86}{2001}{381} [\grqc{0007070}];\\
D.J.H. Chung, E.W. Kolb and A.~Riotto, \emph{Extra dimensions present
  a new flatness problem}, \hepph{0008126};\\
D.~Youm, \emph{Brane world cosmologies with varying speed of light},
\prd{63}{2001}{125011} [\hepth{0101228}];\\
R.R. Caldwell and D.~Langlois, \emph{Shortcuts in the fifth
dimension}, \plb{511}{2001}{129} [\grqc{0103070}].

\bibitem{similar}
D.~Birmingham, \emph{Topological black holes in anti-de~Sitter space},
\cqg{16}{1999}{1197} [\hepth{9808032}];\\
R.-G. Cai, \emph{The Cardy-Verlinde formula and AdS black holes},
\prd{63}{2001}{124018} [\hepth{0102113}];\\
D.~Youm, \emph{The Cardy-Verlinde formula and topological
AdS-Schwarzschild black holes}, \hepth{0105093};
\emph{The Cardy-Verlinde formula and charged topological AdS
  black holes}, \mpla{16}{2001}{1327} [\hepth{0105249}].

\bibitem{TAUB}
A.H.~Taub, \emph{Empty space-times admitting a three parameter group
of motions}, \emph{Annals Math.\ }{\bf 53}{1951}{472};
E.T.~Newman, L.~Tamburino and T.J.~Unti, \jmp{4}{1963}{915}.

\bibitem{ellis}
C.B.~Collins and G.F.R.~Ellis, \emph{Singularities in Bianchi
cosmologies}, \prep{56}{1979}{65}.

\end{thebibliography}
\end{document}